# Fundamental dispersion limit for spectrally bounded On-Off-Keying communication channels

Er'el Granot

*Department of Electrical and Electronics Engineering, Ariel University Center of Samaria, Ariel, Israel*
*erel@ariel.ac.il*

*The fundamental dispersion limit for optical communication based the On-Off-Keying format is calculated. It is shown both analytically and with numerical simulations that an OOK optical sequence, which passes through spectrally narrow non-compensated dispersive channel cannot exceed the limit $\beta_2 B^2 L < \pi^{-1}$, where $\beta_2$, $L$ and $B$ are the dispersion coefficient, the fiber's length and the bit-rate respectively. To the best of our knowledge, this is the first time that such a fundamental limit was formulated. In the literature, only approximation evaluations were developed yielding much smaller limiting values.*

**Introduction.** The On-Off Keying (OOK) is the simplest modulation format, and therefore it is still one of the most ubiquitous methods in optical communications. Evidently, it has many drawbacks in comparison to other technologies[1-4], however, its simplicity makes it the cost effective solution for many applications, mostly, short range and low rates ones, such as Passive Optical Networks (PON's). In these relatively simple and affordable networks, expensive optical elements are scarce, and therefore, it is advisable to build the network without chromatic dispersion fibers (CDF's). Evidently, in currents ordinary PON's it is not affordable to implement coherent detection and fast digital dispersion mitigation. Moreover, in many applications the requirements are very close to the margins of the OOK capabilities. Hence, in these cases, choosing a less noisy amplifier, or adding an optical filter can extend the maximum distance, which can be transmitted by an OOK transponder.

Despite the high prevalence of this method, its fundamental limitations are known only approximately. As far as we know, there was no attempt in the literature to identify the fundamental limit of the problem and only approximate evaluations were made. The criterions, which were used, for which the signal detection is undecidable, were somewhat arbitrary and yielded only rough estimations. As a consequence, there isn't a consensus in the literature regarding the fundamental upper bound. For example, for a Gaussian propagation analysis Agrawal estimated the dispersion limit[5] as $B^2 L \leq c/(2.54 D \lambda^2)$, where $B$ is the Bit-Rate (the reciprocal of the bit period $T = B^{-1}$), $D$ is the dispersion coefficient, $\lambda$ is the carrier's wavelength and $c$ is the speed of light. While similar reasoning led Henry[6] to a little bit higher limit $B^2 L \leq c/(2 D \lambda^2)$, Watts took account of the fact that the bandwidth of the NRZ-OOK signal is 1.4 times the bit rate and found a higher limit [7] $B^2 L \leq c/(1.4 D \lambda^2)$. We will show that these limits are accurate only qualitatively, and that the correct limit in the presence of a spectral filter can be four times higher.

In realistic channels there is always some kind of spectral filtering. Any communication channel is spectrally bounded. Clearly, the presence of a filter with, say, spectral width $\Delta$, will cause signal degradation. In general, the problem is very complex, and is obviously sequence dependent. However, when $\Delta/B$ decreases, the spectral width is narrower and the dispersion tolerance is higher. Obviously, when noise is present in the system then spectrally narrowing the channel may increase the Bit-Error Rate (BER) of the sequence. However, if the noise level is very low, or even can be neglected, then the best performance against dispersion is reached in the narrow channel regime. This regime is important in low noise networks, and is useful in evaluating the fundamental limit of transport in dispersive channels without utilizing dispersion mitigation elements (like DCFs or DSP)[8].

**Theory.** The electromagentic field envelope $A$ is goverend by the linear Schrödinger equation [5](in the meanwhile we ignore higher order dispersion effects) $i \partial A / \partial z = -(\beta_2/2) \partial^2 A / \partial t^2$, where $t = \tau - z/v$ is the time measured with respect to the fiber's time of flight ($\tau$ is the time and $v$ is the light velocity in the fiber).

At $z = 0$ (the beginning of the fiber) an infinite OOK sequence of ideal rectangular pulses is launched, i.e., $A(z=0,t) = \sum_n x_n \operatorname{rect}_{\xi T}(t - nT)$, where $x_n$ is the digitial sequence (either 0 or 1), $\xi$ is a measure of the duty cycle (i.e., it determines the normalized pulses width) and $\operatorname{rect}_{\xi T}(t) \equiv \{1 \ \ for \ \ |t| \leq \xi T/2 \ \ and \ 0 \ \ otherwise\}$. The transfer function of a Gaussain optical filter, with the spectral FWHM $\Delta$, and the dispersive medium can be written $H(f) = \exp\left[-2\ln 2(f/\Delta)^2 + i\beta_2 L(2\pi f)^2/2\right]$, where $\beta_2$ is the dispersion coefficient, and $L$ is the fiber's length. We will later show, that our conclusions are valid for different kinds of filters, which can be considerably (spectrally) sharper, however we strart with a Gaussian filter for it has an exact ananlytical solution[9]. In this case $A(z>0,t) = \sum_n x_n \operatorname{srect}_{\xi T}(t - nT, \Delta, z)$, where

$$\operatorname{srect}_{\xi T}(t, \Delta, z) \equiv \left\{ \operatorname{erfc}\left[(t - \xi T/2)/\sqrt{i 2 \beta_2 z + 2\ln(2)/\pi^2 \Delta^2}\right] - \operatorname{erfc}\left[(t + \xi T/2)/\sqrt{i 2 \beta_2 z + 2\ln(2)/\pi^2 \Delta^2}\right] \right\}/2$$

erfc is the complementary error function[10]. It should be stressed that this expression is equally valid for RZ ($\xi < 1$) and for NRZ ($\xi = 1$) signals, and no approximations were taken.

The amplitude at the center of the bits (where the measurements are taken) at the end of the fiber is (for any bit $m$) $A_m \equiv \sum x_n \, \mathbf{srect}_{\xi T}(T(m-n), \Delta, z)$. Therefore, the "1" and "0" levels are $I(z,"1") = \min_{x_m=1}|A_m|^2$ and $I(z,"0") = \min_{x_m=0}|A_m|^2$ respectively (i.e. the final "1" level is calculated as the minimum of all the initially "1" bits, and the final "0" level is calcualted as the maximum of all the initially "0" bits).

We can therefore define the signal's eye-opening as $\Delta I(z) \equiv I(z,"1") - I(z,"0")$. The first z, for which the eye-opening $\Delta I(z)$ vanishes, is the fiber's maximum length $L_{max}$, beyond which, it is impossible to decode the data. Evidently, this process can provide only the fundamental limit caused by pure dispersion. Any presence of noise will decrease this distance.

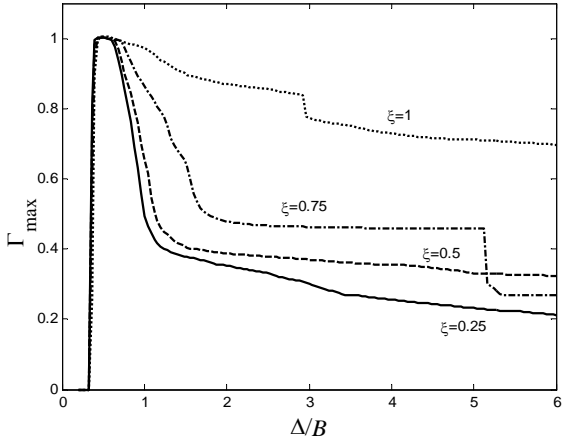

Fig.1. The results of a Monte-Carlo simulation for the normalized maximum distance $\Gamma_{max}$ as a function of the normalized spectral bandwidth $\Delta/B$ for different duty cycles $\xi = 0.25, 0.5, 0.75$ and 1.

**Simulations.** We begin with a Monte-Carlo simulation of a Pseudo Random Bit Sequence of 127 bits (we will later show that the results are independent of the sequence length), which propagates through a Gaussian filter (Later on it will be generalized to other types of filters) with a FWHM of $\Delta$ and a dispersive medium with a coefficient $\beta_2$. For different duty cycles the dimensionless parameter $\Gamma_{max} \equiv \beta_2 \pi B^2 L_{max}$ was calculated as a function of $\Delta/B$. The result are presented in Fig.1. The general tendency is that a decreases in $\xi$ causes a corresponding decrease in $\Gamma_{max}$. This tendency is clear since narrower pulses are prone to disperse faster, due to their wider spectral bandwidth. In all scenarios the normalized distance $\Gamma_{max}$ cannot exceed 1. *This limit is independent of the bit rate, the sequence or the duty cycle.*

**Theoretical Prediction.** In the presence of the optical filter, the most sensitive sequence is the alternating one $1,0,1,0,\ldots 1,0,1\ldots$, which is the digital sequence with the highest frequency. The upper dispersion limit would be determined by the ability to diferenatiate between the "ones" and the "zeros" in this sequence. We will show that this criterion is consistent with exact numrical simulations. As a Fourier series, it can be written

$$A_{01}(z=0,t) = \frac{\xi}{2} + \sum_{n=1}^{\infty} \frac{\sin(\pi n \xi/2)}{\pi n/2} \cos\left(\pi n \frac{t}{T}\right).$$

Then, at the end of the dispersive fiber, the filtered $A_{01}^F(z=L,t)$ signal can be written, after neglecting high harmonics (which is a good approximation due to the low-pass filter):

$$A_{01}^F(L,t) = \frac{\xi}{2} + H\left(\frac{1}{2T}\right)\frac{2}{\pi} \sin\left(\frac{\pi \xi}{2}\right) \cos\left(\frac{\pi t}{T}\right) \exp\left(\frac{i\pi^2 L\beta_2}{2T^2}\right).$$

The intensity at the fiber's end is then $I_{01}^F(L,t) = |A_{01}^F(L,t)|^2$. The sampling time is taken exactly at the center of the symbols, i.e., at $t = mT$, where $m$ is an integer. Hence, the difference between the "one" and the "zero" levels is $\Delta I = 2\pi^{-1}\xi H(B/2) \sin(\pi \xi/2) \cos(\beta_2 L(B\pi)^2/2)$. $\Delta I$ quantifies the signal's eye-opening. When $\Delta I$ vanishes there is no way to decode the signal. That occurs when $B^2 L = (\beta_2 \pi)^{-1}$ and since $\beta_2 = D\lambda^2/2\pi c$, where $D$ is again the dispersion coefficient (in smf28 fiber $D \cong 17\, ps/(nm \cdot Km)$) then $B^2 L_{max} < 2c/D\lambda^2$, which is 2.8 times larger than Watt's evaluation[7], 4 times larger than Henry's [6] and more than 5 times larger than Agrawal's [5]. Clearly, if noise is added then the distance is reduced correspondingly, however, $L_{max}$ is the fundamental upper limit, which cannot be bypassed. It should be noted that this result is independnet of the pulses duty cycle $\xi$. Therefore, RZ and NRZ have the *same* fundamental limit. Obviously, the duty cycle affects the sensitivity of the pulse to dispersion and filtering, however, in the spectrally narrow channel regime all the pulses have the same limit.

**Higher Order Dispersion.** Similarly, if the second order dispersion is cancelled, i.e., $\beta_2 = 0$, then the next significant term is the third one, i.e., $G(f) = \exp(i\beta_3 L(2\pi f)^3/6)$. This is the case in certain DSF, or a dispersive channel, which was mitigated by DCF.

From exactly the same reasoning that led to $\Gamma_{max}$ we can deduce, that the maximum distance for medium, which is governed by a third-order dispersion, must obey $\Gamma_{max}^{(3)} \equiv \beta_3 \pi^2 B^3 L_{max}/3 < 1$. This maximum distance is almost 9 times larger than Ref.[5] (in this reference $B^3 L \leq (2/3)^{3/2}/16\beta_3$). In Fig.2 we plot a Monte-Carlo simulation result of the normalized maximum distance $\Gamma_{max}^{(3)}$ as a function of $\Delta/B$. As can be seen, the normalized maximum distance cannot exceed the value 1. This result can be generalized to any dispersion order. For an $n$th-order dispersion medium, i.e., $G(f) = \exp(i\beta_n L(2\pi f)^n/n!)$, the limit is $B^n L_{max} < n!/\beta_n \pi^{n-1}$.

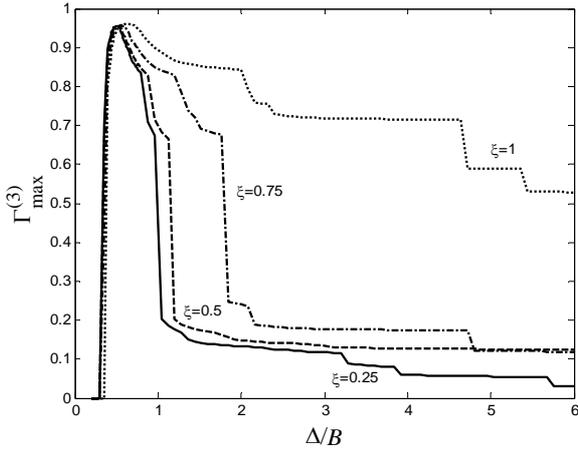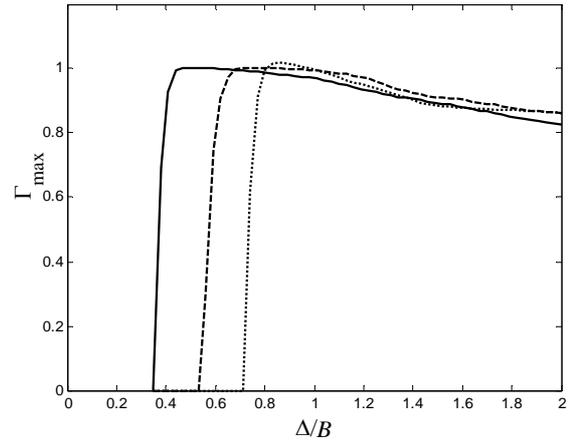

Fig.2: A Monte-Carlo simulation result for the normalized maximum distance $\Gamma_{max}^{(3)}$ as a function of the normalized spectral bandwidth $\Delta/B$ for different duty cycles $\xi = 0.25, 0.5, 0.75$ and 1.

Fig.3: Like Fig.1 but for different $n$ (filter's order). The solid curve, the dashed one, and the dotted one stand for $n=2,4,8$ respectively. The number of bits was 8191 and $\xi = 1$ (NRZ).

**Filters with different shapes.** It is clear from the derivation of the fundamental limit, that the limit is practically independent of the filter's shape. In fact, the criterion reasoning is even more valid for a sharper filters. Take, for example, filters, with the generic shape (supergaussians): $H_n(f) = \exp\left[-2^{n-1}\ln 2 (f/\Delta)^n\right]$, where $n$ is the filter's order. Clearly, $n = 2$ stands for a Gaussian filter and $n \to \infty$ stands for an ideal LP rectangular filter. The results of a Monte-Carlo simulation for different kind of filters are presented in Fig.3 for a PRBS series of $2^{13} - 1 = 8191$ bits. As can be seen from Fig.3, the maximum distance is practically independent of the filter's shape. The bandwidth, for which maximum distance is achieved, depends on the shape of the filter. For sharp filters the bandwidth cannot be considerably narrower than the data rate, while for moderate filters, due to their long spectral tails, the filter FWHM can be considerably narrower than the bandwidth.

It should be noted that since Gaussian and Nyquist pulses, are actually spectrally filtered delta functions, *they are also subject to the same limit.*

**Conclusions and Summary.** Optical data transmission based on OOK is subject to the fundamental limit (which is considerably larger than what was written in the literature) $B^2 L_{max} < 2c/D\lambda^2$, which is independent of the sequence, the protocol, the duty cycle or the spectral width and shape of the optical channel. Similar conclusions are also generalized to higher orders of dispersion.